\documentclass[%
 reprint,
 amsmath,amssymb,
 aps,
prb,
]{revtex4-2}

\usepackage{graphicx}% Include figure files
\usepackage{dcolumn}% Align table columns on decimal point
\usepackage{bm}% bold math
\begin{document}

\preprint{APS/123-QED}

\title{Impurity metallic conduction below the critical concentration of Metal-Insulator Transition in Fe$ _{1-x} $Co$ _{x} $Si  }% Force line breaks with \\
%\thanks{A footnote to the article title}%

\author{M. Krishnan}
\email{Corresponding author : krishnanmlm@gmail.com}
 %\altaffiliation[Also at ]{Physics Department, XYZ University.}%Lines break automatically or can be forced with \\
 \affiliation{%
 UGC-DAE Consortium for Scientific Research, University Campus, Khandwa Road, Indore 452 001, India.}%
%\affiliation{
% Department of Physics, Kyoto University, Kyoto 606-8502, Japan
%}%

\author{V. Ganesan}
% \homepage{http://www.Second.institution.edu/~Charlie.Author}
\affiliation{
 School of Advanced Science and Languages (SASL), VIT Bhopal University, (MP) 466114, India.
}%

%\collaboration{MUSO Collaboration}%\noaffiliation

%\author{Charlie Author}
% \homepage{http://www.Second.institution.edu/~Charlie.Author}
%\affiliation{

%\affiliation{Third institution, the second for Charlie Author}%
%\author{Delta Author}
%\affiliation{%

%\collaboration{CLEO Collaboration}%\noaffiliation

\date{\today}% It is always \today, today,
             %  but any date may be explicitly specified

\begin{abstract}

Analysis on very detailed measurements of resistivity ($ \rho $) and thermoelectric power (S) of magnetic impurity (Co) substituted iron silicide (FeSi) has been presented in this report. The impurity valence electrons of Co dominate the whole physical properties at low temperatures below 35 K, below the critical concentration x$_{c}$. The negative thermopower and the positive slope in the resistivity at low temperatures are exotic and show that the system is not entirely insulator below the critical concentration of MIT (x$_{c}$). So, due to the external impurity electrons, the system's magnetic ground state could change considerably compared to the parent compound FeSi. This report may help unveil the exotic nature of the ground state in the semi-metallic regime between x = 0 to x = 0.02. We have also explained the electrical and thermal transport properties using the two-band model.    

\end{abstract}

%\keywords{Suggested keywords}%Use showkeys class option if keyword
                              %display desired
\maketitle

%\tableofcontents

\section{\label{sec:level1}INTRODUCTION}

Semiconductor contribution is enormous in electronics industries, including spintronics, and knowledge of transport properties is necessary. It gives a basic model to theorize transport properties at absolute zero. The scientific community has always had a keen interest in semiconductors due to their unusual transport behavior by external parameters. The physical properties of semiconductors at low temperatures are exotic and anomalous due to the presence of finite impurities. So it is necessary to study impurity effects on transport properties at low temperatures. Metal-Insulator Transition (MIT) is one fascinating physical phenomenon that has been studied over several decades. Physical properties around MIT helps to understand how the transport properties are greatly affected by external parameters such as magnetic field, impurities, pressure, etc. Classical semiconductors such as Silicon (Si) or Germanium (Ge) have been rigorously studied under the above external conditions over the past several decades, and their physical properties are well analyzed with various impurities such as Al, P, As \cite{Rosenbaum1994} \cite{Mostafa2017} \cite{Nakamura2003}\cite{Markevich2004}. FeSi is a hybridized semiconductor whose physical properties are exotic at low temperatures. Alloying transition metal iron and metalloid silicon make a unique material whose behavior is neither metal nor completely semiconductor. FeSi crystallizes B20 cubic form with space group P2$ _{1} $3. Each unit cell of FeSi contains eight atoms. Exotic properties of the ground state are connected with the non-centrosymmetric arrangement of Fe and Si atoms in unit cell \cite{pauling1948}. At room temperatures, FeSi behaves like metallic, whereas while cooling to low temperatures, it changes to activated behavior due to a gap opening around the Fermi level \cite{Hunt1994} \citep{sales1994}. FeSi is paramagnetic metal at high temperatures above $\sim$400 K, but when decreasing temperature, resistivity increases and saturates below 10 K. Activation energy of FeSi is 50 meV at low temperatures \cite{Jaccarino1967}\cite{sales1994}.

Researchers suggest that the appearance of non-magnetic singlet ground state at low temperatures is of Kondo type but still debate going on \cite{Coleman2007}. The ground state of this system is distinct in the sense that electronic properties cannot be explained with the classical semiconductor model. The impurities make the ground state of the system very sensitive. Cobalt (Co) has one extra electron as compared to the Iron in the $\textit{3d}$-orbital, and adding Co in the FeSi makes the system becomes electron-dominated. The Co-doped FeSi shows MIT along with itinerant ferromagnetic behavior above x$ _{c} $ = 0.02, hellimagnetic behavior above x = 0.05 \cite{Chernikov1997} \cite{Manyala2000}  \cite{Guevara2004} \cite{Mazurenko2010} and a non-Fermi liquid can be derived with Mn as impurities \cite{Manyala2008}. Also, The 20 percentage Co-doped FeSi shows skyrmion magnetic structure, despite the strong site disorder due to doping \cite{Munzer2010}.  Ge has a similar electronic configuration as Si and, by substituting in Si site makes ferromagnetic metal \cite{Yeo2003}. FeSi and CrSi are paramagnetic metal \cite{Yadam2016}, MnSi is itinerant ferromagnet and CoSi is diamagnetic metal \cite{Manyala2008}. All these varieties of ground states can be achieved with the transition from semiconducting to metallic nature at low temperatures with the various dopant. So by understanding the formation of impurity states around the Fermi level will help to gain knowledge on ground-state properties and impurity effects at low temperatures.

\section{\label{sec:level1}EXPERIMENTAL METHODS}

Cobalt doped FeSi samples were prepared by using the arc melting technique. High purity constituent elements such as Fe, Co, and Si were loaded in a double-walled arc melting chamber, and the rod-shaped samples were obtained by melting the constituent elements in an argon environment to avoid oxidation at higher temperatures. Ti ball were used as a gettering element to observe residual impurities. The resultant rod-shaped samples were sealed in a quartz tube which was at a high vacuum better than 10$ ^{-6} $ mbar pressure. The sealed quartz tubes were loaded in the furnace for annealing at 1000º C for a week. The purpose of annealing was to remove the strain from sudden quenching in the preparation of arc melting and improve better homogeneity.

Electrical transport property measurement was carried out in 2K/14T PPMS (QD USA) using the dc four-probe method. Moreover, Thermoelectric power measurement had been done by using CCR homemade differential dc sandwich method \cite{SharathChandra2008b}.

\section{\label{sec:level1}RESULTS AND DISCUSSIONS }

\subsection{\label{sec:level2}XRD analysis of Fe$ _{1-x} $Co$ _{x} $Si}

\begin{figure}
	\includegraphics[width=10cm,height=10cm]{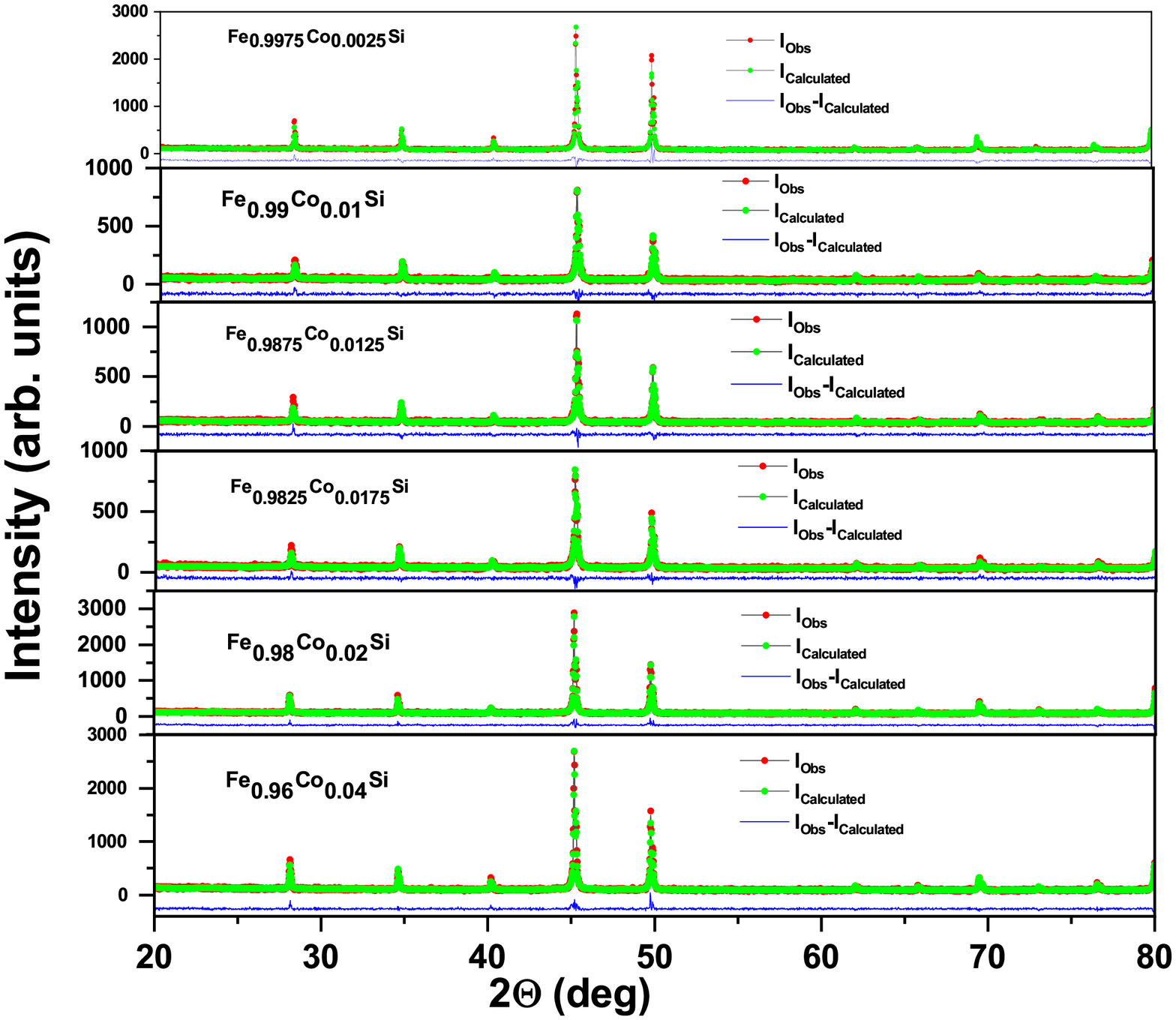}% Here is how to import EPS art
	\caption{\label{fsc_xrd} XRD of Fe$ _{1-x} $Co$ _{x} $Si (x = 0.0025, x = 0.01, 0.0125, 0.0175, 0.02, 0.04).}
\end{figure}

The Bruker D8 X-ray diffractometer ($ Cu-K \alpha $ radiation) was used to confirm the phase purity of the samples. Fullprof software was used to refine the powder diffraction data. The diffraction data shown in Fig.~\ref{fsc_xrd} with Reitveld refinement. The lattice parameters were extracted and tabulated in table \ref{tfsc_xrd}. The lattice parameters are not varying considerably with Co concentration. The absence of impurity peak in the system and the single-phase nature of samples implies that the Co atoms successfully replaced Fe atoms.

\begin{table}[!htb]
  \centering
  \begin{tabular}{cc}
  \hline
  Fe$ _{1-x} $Co$ _{x} $Si (x) & Lattice parameter (\AA)\\
  \hline
  0.0025 &  4.4884 $\pm$ 0.0003\\
  0.01 &   4.4854 $\pm$ 0.0003\\
  0.0125 & 4.4811 $\pm$ 0.0007\\
  0.0175 & 4.4857 $\pm$ 0.00001\\
  0.02 &   4.4852 $\pm$ 0.0006\\
  0.04 &   4.4845 $\pm$ 0.000008\\
 \hline
  \end{tabular}
  \caption{Lattice parameter of Fe$_{1-x}$Co$_{x}$Si}
  \label{tfsc_xrd}
\end{table}

\subsection{\label{sec:level2}Electrical resistivity of Fe$ _{1-x} $Co$ _{x} $Si}

\begin{figure}
	\includegraphics[width=9cm,height=7cm]{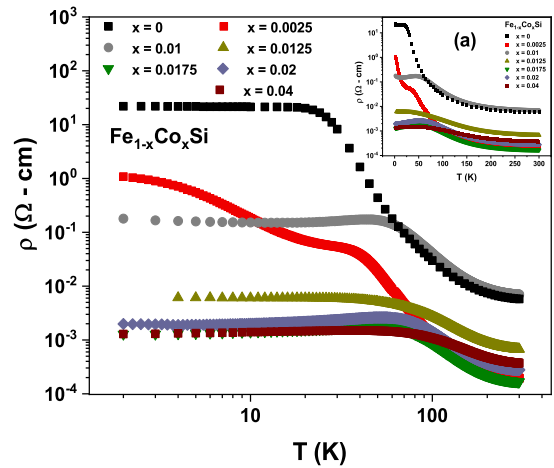}% Here is how to import EPS art
	\caption{\label{FSC_RT} Electrical resistivity of Fe$ _{1-x} $Co$ _{x} $Si (x = 0.0025, 0.01, 0.0125, 0.0175, 0.02, 0.04). Inset: (a) Resistivity of Fe$ _{1-x} $Co$ _{x} $Si in linear versus log plot.}
\end{figure}

Figure \ref{FSC_RT} demonstrates the electrical resistivity of Fe$ _{1-x} $Co$ _{x} $Si for different level of cobalt substitutions (x = 0.0025 to 0.04) at the temperature interval from 2 to 300 K. The Figure \ref{FSC_RT} inset (a) shows the electrical resistivity of Fe$ _{1-x} $Co$ _{x} $Si in  linear versus log plot. The resistivity of FeSi at 300 K is 5.7 milli-$\Omega$-cm, which is indicative of metallic nature whereas, as we cool down the system from 300 to 2 K, the resistivity increases by four orders of magnitude due to activation of intrinsic majority electrons. At low temperatures below 30 K, the resistivity saturates due to hopping conduction of extrinsic minority holes \cite{Krishnan2020}. The overall resistivity of Co-doped FeSi decreases with increasing Co concentration. The resistivity of x = 0.0025 changes by four orders of magnitude in the whole temperature range. The resistivity below 30 K shows slight slope changes due to the change in nature of carrier conduction (from p-type to n-type). Below 20 K, the resistivity shows logarithmic increasing behavior.
Furthermore, this low-temperature nature is different from the behavior of FeSi. This might be due to the electron-hole recombination; as a result, the number carrier concentration decreases. Hence, the resistivity shows insulating behavior. The resistivity below 40 K shows dramatic changes at higher concentrations above x = 0.0025, which we will see in detail. In x = 0.01, the electrical resistivity shows a positive slope or a hump below/at 40 K, indicating metallic behavior. This metallic conduction is due to the n-type carriers from the external doping.

Further, the resistivity shows a negative slope below 15 K. This implies a competition between the impurity holes from FeSi and the impurity electrons from the Co external doping. With further increasing the Co doping concentration, the positive slope continues, and the low-temperature negative slope disappears, implying the domination of impurity electron conduction. The appearance of the positive slope in all the Co doping concentrations implies that the metallic/impurity-band conduction starts even at a very low concentration, below x$ _{c} $. The temperature at which the slope changes also move towards the higher temperature. This indicates that the metallic conduction dominates over the semiconducting nature with increasing Co concentration.

\begin{figure}
	\includegraphics[width=9cm,height=7cm]{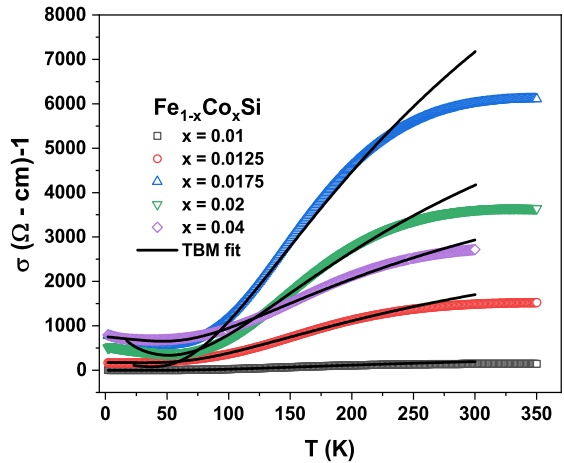}% Here is how to import EPS art
	\caption{\label{RT_TBM} Electrical conductivity of Fe$ _{1-x} $Co$ _{x} $Si with TBM fit.}
\end{figure}
    
\begin{eqnarray}
\sigma (T) = \dfrac{1}{\rho_{01} + AT} + \sigma_{02}exp(-\dfrac{E_{g\rho}}{T})
\label{TBMREqt}
\end{eqnarray}

\begin{table}[!htb]
  \centering
  \begin{tabular}{ccccc}
  \hline
  Concentration & $\frac{1}{\sigma _{01}} $ & \textit{A}$ _{\rho} $  & $\frac{1}{\sigma _{02}} $  & $E_{g\rho}$ \\
(x) & ($\Omega~cm$) &($\Omega~cm$ $K^{-1}$)&  ($\Omega~cm$) & (K) \\
  \hline
0.01 & 9.1 & 2480 & 0.0018 & 315 \\
0.0125 & 173.6 & 3.34$\times 10^{-6}$ & 0.000243 & 292 \\
0.0175 & 2102.16 & 3.74$\times 10^{-5}$ & 5.4$\times 10^{-5}$ & 285 \\
0.02 & 1924.1 & 5.68$\times 10^{-5}$ & 9.6$\times 10^{-5}$ & 279 \\
0.04 & 757.57 & 5.5$\times 10^{-6}$ & 1.54$\times 10^{-4}$ & 275 \\
  \hline
  \end{tabular}
  \caption{Coefficients derived from TBM fit for the resistivity of Fe$ _{1-x} $Co$ _{x} $Si.}
  \label{tfsc_tbm}
\end{table}
 
 From the observations of resistivity, it is clear that two kinds of bands dominate the resistivity. In the temperature interval between 50 - 200 K, the resistivity increases exponentially due to the activation of the majority of intrinsic electrons. The temperature below 40 K, the resistivity, shows a positive slope, which shows the metallic conduction, which is due to the impurity band from the Co substitution. So, The high-temperature band behaves as a semiconducting band, whereas the low temperature behaves as a metallic band. Hence, the Two-Band Model (TBM) can be used to understand the effect of the semiconducting as well as metallic impurity band and to calculate the activation energies of Fe$ _{1-x} $Co$ _{x} $Si \cite{Krishnan2020} \cite{SharathChandra2008iop}.

 The first-term in the right hand side of the Equation.~\ref{TBMREqt} is for the metallic band ($\sigma_{1} = (\sigma_{01} + A_{\rho}T)^{-1}$) responsible for impurity conduction, and the second term is for the semiconducting band $\sigma_{2} = \sigma_{02}exp(-\dfrac{E_{g\rho}}{T}$) responsible for the activated conduction. $E _{g\rho}$ is the activation energy and $\frac{1}{\sigma_{01}}$, $\frac{1}{\sigma_{02}}$ are the background resistivity due to scattering from the metallic and the semiconducting band respectively \cite{Krishnan2020} \cite{SharathChandra2008iop}. The coefficients derived from Equation.~\ref{TBMREqt} tabulated in Table.~\ref{tfsc_tbm}. The Coefficient $A_{\rho}$ from the metallic term varies non-monotonically with Co concentration. The activation energy (E$_{g\rho}$) decreases with increasing Co concentration.

\subsection{Thermopower of Co substituted FeSi}

\begin{figure}
\centering
	\includegraphics[width=9cm,height=7cm]{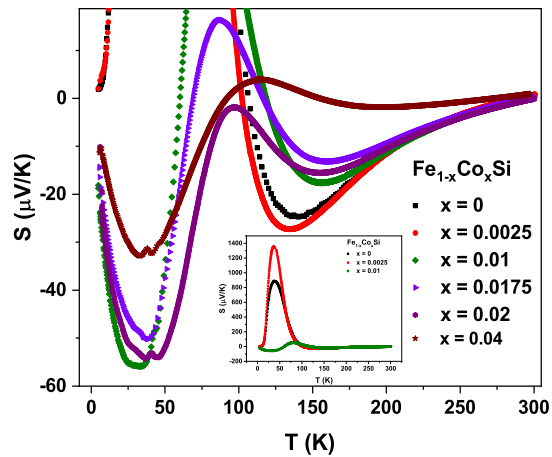}% Here is how to import EPS art
	\caption{\label{fsc_tep1} Thermopower of Co substituted FeSi from x = 0 to x= 0.04.}
\end{figure}

The thermopower data of Fe$_{1-x} $Co$_{x}$Si from x = 0 to x = 0.04 shown in Fig.~\ref{fsc_tep1}. The measurement is done on the temperature interval between 5 - 300 K. The thermopower of parent compound FeSi shows positive values at room temperature (S = +0.4 $ \mu $V/K), and this is due to the dominance of high mobility holes. These holes are extrinsic, which is created by the slight deficiency of silicon during sample preparation \cite{Degiorgi1999}. The decrease in temperature makes the thermopower negative and reaches a maximum in the negative direction around 110 K. This negative thermopower is due to the activation of electrons across the semiconducting band. The thermopower suddenly goes in a positive direction with further decreasing temperature and reaches a positive maximum at 40 K, implying that the carrier conduction changes from electrons to holes. This positive thermopower is due to extrinsic holes. The thermopower goes to zero temperature goes to near zero. From the nature of thermopower of FeSi, two kinds of carriers dominate in the conduction process, namely the minority holes conducting at a high temperature above 300 K as well as at low temperature below 60 K and the majority carriers dominating in the mid-range of temperature interval between 60 - 290 K. 

 The substitution of cobalt makes dramatic changes in the low-temperature thermopower of FeSi. The thermopower at x = 0.0025 shows an enhancement at low temperatures compared to the parent FeSi. The value of thermopower for x = 0.0025 is +1350 $\mu V/K$ at 36 K, which is 460 $\mu V/K$ higher than the parent FeSi. The thermopower is also enhanced at 130 K in the negative direction. It is interesting to note that the dopant thermopower is higher than the parent compound. This is as follows; cobalt has one extra electron as compared to Fe. The substitution of cobalt adds electron-like density of states near the conduction band, and these impurity electrons make recombination with holes. As a result, the number of extrinsic holes and intrinsic electrons decreases. So, the thermopower increases, because it is inversely proportional to the carrier concentration ($S \sim \gamma T/ne$) \cite{SharathChandra2008prb}. The increase in cobalt concentration makes the thermopower negative at low temperatures below 40 K at the concentration above x = 0.0025. The negative thermopower is due to the domination of the impurity electrons. The low-temperature interval of negative thermopower between 5 to 50 K increases with increasing Co concentrations. At the same time, positive thermopower is drastically suppressed with Co concentration. The peak in the negative thermopower at 125 K decreases with increasing Co concentration. At x = 0.04, the thermopower is fully negative except in the narrow interval between 90 to 150 K with a small positive thermopower. At room temperature, it does not vary considerably.

\begin{figure}
\centering
	\includegraphics[width=9cm,height=7cm]{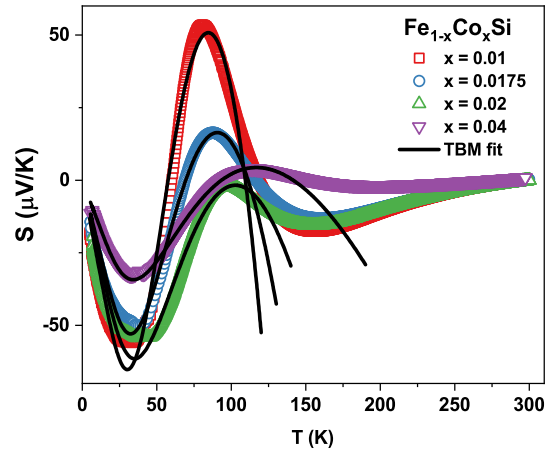}% Here is how to import EPS art
	\caption{\label{fsc_tep2} Thermopower of Co substituted FeSi from x = 0 to x= 0.04 with two band model (TBM) fit.}
\end{figure}

The nature of thermopower in Co-doped FeSi shows that the two kinds of bands dominate: 1) metallic band extrinsic carriers and 2) the semiconducting band due to the intrinsic carriers. The Two-Band Model (TBM) helps find the effect of impurity carriers and the intrinsic carrier contributions. A simplified version of TBM used here \cite{Xin1993} \cite{Krishnan2020}:

 \begin{eqnarray}
 S = \dfrac{S_{1}\sigma_{1}+S _{2}\sigma _{2}  }{\sigma_{1}+\sigma_{2}}
 \label{TBMTEq1}
 \end{eqnarray}
 
 Where S$_{1}$ = $\pi$$^{2}$k$_{B}$$^{2}$T/2eE$_{F}$ represents diffusion thermopower and S$_{2}$ = c + (k$_{B}$/e)(E$_{gS}$)/2k$_{B}$T represents semiconducting contribution of TEP.

 \begin{table}[!htb]
  \centering
  \begin{tabular}{cc}
  \hline
  Concentration & E$_{gS} $ \\
(x) & (K) \\
  \hline
0.01 & 157.7 \\
0.0175 & 151.9 \\
0.02 & 141.5  \\
0.04 & 128.5  \\
  \hline
  \end{tabular}
  \caption{Energy gap derived from TBM fit for the thermopower of Fe$ _{1-x} $Co$ _{x} $Si.}
  \label{fsc_ttbm}
\end{table}

 Fig.~\ref{fsc_tep2} shows the thermopower from x = 0 to x = 0.04 with TBM fit. The activation energy (E$ _{gS} $) derived by using two band model listed in Table.~\ref{fsc_ttbm}. The E$ _{gS} $ decreases with increasing Co concentration monotonically, which implies the closure of the gap.

 \

 \section{\label{sec:level1}Conclusion }
  We investigated the transport and thermoelectric properties of Co-doped FeSi. We concentrated more on the physical properties, which are below the critical concentration (x$_{c}$ = 2 \%) of MIT and below x$_{c}$, where the system is non-magnetic. As far as our knowledge, we are the first to report the detailed thermoelectric properties of Fe$ _{1-x} $Co$ _{x} $Si below x$_{c}$ at low temperatures. The overall resistivity decreases with increasing Co concentration, and we used the two-band model to get an insight into the impurity effects. The activation energy decreases with increasing Co concentration, implying that the impurity states form near the conduction band. The positive thermopower around 40 K dramatically increases at x= 0.0025 and then decreases with increasing concentration. The thermopower shows a negative below 35 K from x = 0.01; implies the impurity electron carrier domination at low temperatures. The overall activation energy derived by using the two-band model in thermopower decreases with increasing Co concentration.    
 
\begin{acknowledgments}
 
 The Authors thank Director and Centre-Director, UGC-DAE CSR, Indore and Dr. R. Venkatesh for their support. Mr. P. Saravanan is thanked for his help in cryogenics as well as in automation of TEP set up, inmates of LTL and Cryogenic lab for their technical support. DST India is thanked for their initial support in setting up of a 14T PPMS system. This work was done when the authors M. K and V.G were at UGC-DAE-CSR, Indore. The author M. K currently working at Kyoto University, Japan, under the supervision of Prof. Yoshiteru Maeno, JSPS grant No. JP17H06136.

\end{acknowledgments}

\bibliography{FSC}% Produces the bibliography via BibTeX.

\end{document}